\documentclass[onecolumn,draftclsnofoot ,12pt]{IEEEtran}

\usepackage{times}
\usepackage{pstool}
\usepackage{subfigure}
\usepackage{multirow}
\usepackage{enumerate}
\usepackage{stfloats} 
\usepackage[table]{xcolor}
\usepackage{amsmath,graphicx}
\usepackage{algorithm, algorithmic}
\usepackage{epsfig, graphics, color}
\usepackage{amsmath, amssymb, amsfonts, amsbsy, mathrsfs, bm}
\usepackage{arydshln, multirow}
\usepackage{cases, enumerate, fancyvrb}
\usepackage{theorem}
\usepackage[compress]{cite}
\usepackage{psfrag}
\usepackage[normalem]{ulem}
\usepackage{verbatim}
\usepackage{subfigure}

\DeclareMathAlphabet{\mathsfsl}{OT1}{cmss}{m}{sl}

\newtheorem{theorem}{\bf Theorem}
\newtheorem{proposition}{\bf Proposition}
\newtheorem{lemma}{\bf Lemma}

\newtheorem{assumption}{\bf Assumption}

\newcommand{\bg}{\mathbf{g}}

\newcommand{\bp}{\mathbf{p}}
\newcommand{\bq}{\mathbf{q}}

\newcommand{\bs}{\mathbf{s}}
\newcommand{\bt}{\mathbf{t}}

\newcommand{\bx}{\mathbf{x}}
\newcommand{\by}{\mathbf{y}}

\newcommand{\V}{\textrm{Vec}}

\newcommand{\td}{\tilde{d}}
\newcommand{\tn}{\tilde{n}}
\newcommand{\tth}{\tilde{\theta}}
\newcommand{\ts}{\tilde{s}}
\newcommand{\tilt}{\tilde{t}}

\newcommand{\bars}{\bar{s}}
\newcommand{\bart}{\bar{t}}
\newcommand{\barth}{\bar{\theta}}

\newcommand{\bA}{\mathbf{A}}
\newcommand{\bB}{\mathbf{B}}

\newcommand{\bU}{\mathbf{U}}
\newcommand{\bV}{\mathbf{V}}

\newcommand{\dd}{\mathrm{d}}

\newcommand{\stE}{\mathcal{E}}

\newcommand{\stS}{\mathcal{S}}

\newcommand{\stT}{\mathcal{T}}

\newcommand{\R}{\mathbb{R}}

\newcommand{\E}{\mathbb{E}}

\newcommand{\norm}[1]{{\left\lVert #1 \right\rVert}}

\newcommand{\condp}[2]{p \left(\left. #1 \right| #2 \right)}
\newcommand{\Qfun}[2]{Q\left(\left. #1 \right| #2 \right)}

\newcommand{\bigangle}[1]{\langle#1\rangle}

\graphicspath{{Figures/}}

\begin{document}

\title{Distributed Localization of a RF target in NLOS Environments}

\author{
\IEEEauthorblockN{Wenjie\ Xu, Fran\c{c}ois\ Quitin, Mei\ Leng, Wee\ Peng\ Tay and Sirajudeen\ G.\ Razul}
}

\definecolor{gray90}{gray}{0.9}

\maketitle
\thispagestyle{empty}
\pagestyle{empty}

\long\def\symbolfootnote[#1]#2{\begingroup%
\def\thefootnote{\fnsymbol{footnote}}\footnote[#1]{#2}\endgroup}
\symbolfootnote[0]{\hrulefill \\ 
Wenjie\ Xu, Fran\c{c}ois\ Quitin, Mei\ Leng and Wee\ Peng\ Tay are with the School of Electrical and Electronic Engineering, Nanyang Technological University, Singapore. Sirajudeen\ G.\ Razul is with Temasek Laboratories at Nanyang Technological University, Singapore. 
(\{{wjxu}, {fquitin}, {lengmei}, {wptay}, {esirajudeen}\}@ntu.edu.sg)\\}

\begin{abstract}
We propose a novel distributed expectation maximization (EM) method for non-cooperative RF device localization using a wireless sensor network. We consider the scenario where few or no sensors receive line-of-sight signals from the target. In the case of non-line-of-sight signals, the signal path consists of a single reflection between the transmitter and receiver. Each sensor is able to measure the time difference of arrival of the target's signal with respect to a reference sensor, as well as the angle of arrival of the target's signal. We derive a distributed EM algorithm where each node makes use of its local information to compute summary statistics, and then shares these statistics with its neighbors to improve its estimate of the target localization. Since all the measurements need not be centralized at a single location, the spectrum usage can be significantly reduced. The distributed algorithm also allows for increased robustness of the sensor network in the case of node failures. We show that our distributed algorithm converges, and simulation results suggest that our method achieves an accuracy close to the centralized EM algorithm. We apply the distributed EM algorithm to a set of experimental measurements with a network of four nodes, which confirm that the algorithm is able to localize a RF target in a realistic non-line-of-sight scenario. 
\end{abstract}
\begin{keywords}
Target localization, Expectation maximization, Decentralized algorithm, Wireless sensor network
\end{keywords}
\section{Introduction}
\label{sec:intro}

% Wireless sensor networks for localization
A wireless sensor network (WSN) consists of a large number of distributed devices that have limited sensing, communication and processing capabilities. RF target localization is an important application of WSNs in which multiple sensors collect and cooperatively process the location information gathered from the wireless signal transmitted by the target.  
% Examples of applications for WSN localization
There are many fields in which WSN localization is used. One example is the accurate localization of mobile phones in search and rescue operations, which has the potential to significantly reduce emergency response time. Another example is tracking of wildlife in large areas, which can be challenging because of the large areas to be monitored. A more commercial application is the localization and tracking of customers in a shopping area, or the precise localization of social network users in urban environment. The latter can be challenging because the line-of-sight between the RF target and sensors is often obstructed in cluttered urban environments. \\
% Cooperative=>TOA, Non-cooperative=>TDOA,AOA
When the target is cooperative, and its waveform signature is known, it is possible to use time-of-arrival (TOA) techniques to localize the target, provided the target and the sensors can be synchronized. However, in the case of non-cooperative RF targets, one must resort to time-difference-of-arrival (TDOA) or angle-of-arrival (AOA) techniques. 
% LOS vs NLOS
These techniques have been widely investigated in the context of line-of-sight (LOS) scenarios, where a LOS exist between the target and the different sensing nodes. In practical cases, however, it is rare to have LOS signals, especially in urban environments. 
% Centralized vs distributed
Another problem is that traditional localization algorithms are centralized, and they require all sensor data to be transmitted to a fusion sensor. In practice, sensors are often randomly scattered in the environment, and each sensor has limited energy storage. If a random sensor is selected to be the fusion sensor, it may quickly deplete its energy reserve due to the high processing costs involved. Such a network is also vulnerable to the failure of the fusion sensor. Distributed algorithms, on the other hand, have drawn a lot of attention due to the fact that they can distribute the processing load among sensors by leveraging local data transmission and local processing. 
% This paper
In this work, we investigate the target localization problem when we have predominantly non-line-of-sight (NLOS) signals. We model the NLOS signals using a single-bounce path between target and sensor, without knowing the location of the scatterer. As the target is non-cooperating, we consider that only TDOA and AOA information is available at the sensors, and we investigate the use of distributed algorithms to perform the target localization. 

\subsection{Main Contributions} 
The contributions of this paper can be summarized as follows: 
\begin{itemize}
 \item We propose a distributed generalized expectation maximization (EM) algorithm for non-cooperative RF target localization in NLOS environments. The distributed EM algorithm utilizes TDOA and AOA information at each sensor node, and treats the orientations of the scatterers from which signals are bounced off as unobserved latent variables. The locations of the scatterers are assumed to be unknown. We provide a proof to show that, under some sufficient conditions, the proposed distributed algorithm converges to a local optimum of the likelihood function. 
 \item We provide simulation results to show the performance of the distributed EM algorithm, and compare it with a centralized EM algorithm that serves as a benchmark. We analyze the performances of the results for different TDOA and AOA noise values. 
 \item We perform an experimental measurement campaign and apply the distributed EM algorithm on a set of experimental results in which we have 1 LOS and 3 NLOS nodes to the target. Our results show that the distributed EM algorithm performs well when applied on this realistic scenario, and achieves localization accuracies below 15~m for 90\% of the measurements, where the TDOA measurements are up to 160~m. 
\end{itemize}

%Wang:TVT:2013,
\subsection{Related Works} 
Localization techniques based on wireless TDOA have been extensively investigated, ranging from a few decades back \cite{Knapp:TASSP:1976,Jacovitti:TSP:1993,Chan:TSP:1994} to the present \cite{Sathyan:AES:2006,Ho:SP:2007,Mao:WSN:2007,Wang:TWC:2011,Choi:IET_SP:2013}. Approaches robust to NLOS errors have been proposed in \cite{Cong2001, Ekambaram2013, Yi2013}. These approaches assume that a sufficient number of sensors receive the LOS signals, and NLOS errors are filtered out by making use of either LOS measurements or a predicted target position. Also of interest is \cite{Algeier2008} where the measured MIMO signal is compared with a ray-tracing-generated database to determine the transmitter's location. However, the latter requires to have a precise map of the environment, which might be impractical in many applications. In \cite{Yin2013}, the authors use TOA and propose to estimate the measurement probability density function (pdf) through an iterative process. Once the measurement pdf is known, one is able to retrieve the target location. In \cite{Yin2014}, the TOA is modeled as a two-mode mixture distribution (even though the underlying process may be different) whose parameters are estimated through different techniques. However, both of these papers assume that different measurements at a particular location yields i.i.d. measurements. In real scenarios, the TOA/TDOA measurement error of static nodes will be dominated by multipath, and will not be i.i.d. Moreover, \cite{Yin2013, Yin2014} assumes that the TOA measurement pdf remains identical for a given environment, which might prove to be a somewhat unrealistic assumption in practical scenarios. 

% NLOS signals, and scatterer orientation as intro to EM algorithm
NLOS signals can be modeled using a ray-tracing model that uses the TOA of a single bounce path between the target and sensor, as well as the corresponding AOA and angle-of-departure (AOD)\cite{Li2007,Miao2007,Seow2008}. However, since we assume that our target is non-cooperative, sensors have no access to the AOD information. Furthermore, we do not assume that sensors have knowledge of the target signal structure, so only TDOA and AOA information are available at the sensors. In this formulation, the orientation angles of the scatterers are treated as latent variables, therefore we appeal to the EM algorithm \cite{Dempster1977} to estimate the target position.  A distributed EM algorithm where a message has to cycle across the entire network through a predefined sensor sequence has been proposed in \cite{Nowak2003}, which makes it susceptible to sensor failures. Another distributed EM algorithm has been recently proposed in \cite{Morral2012, Cappe2009}, where the local likelihoods at each sensor belong to the exponential family. The system model in our problem unfortunately does not belong to this class of distributions, and we have to develop a new distributed EM algorithm based on \cite{Fessler1994} for target localization in NLOS environments. The EM algorithm uses an alternating maximization at the M-step performed at each sensor. We show that under certain technical conditions, our distributed EM algorithm allows the target position estimates at all sensors to converge to the same estimate, which is a local maximizer of the likelihood function. 

The rest of this paper is organized as follows. In Section \ref{sec:system_model}, we present our system model. We briefly review the centralized EM method for target localization, and then propose a distributed EM algorithm in Section \ref{sec:em_algo}. In Section \ref{sec:conv_anal}, we provide a convergence analysis for our distributed EM algorithm, and we verify its performance through simulations in Section \ref{sec:simul}. We evaluate the performance of our distributed EM algorithm in a real scenario by using measurements collected with our USRP-software defined radio platrform in Section \ref{sec:exper}. Finally, we conclude in Section \ref{sec:conclu}.
%We implement our distributed EM algorithm on a USRP-software defined radio platform to evaluate our algorithm in a realistic scenario in Section \ref{sec:exper}. Finally, we conclude in Section \ref{sec:conclu}.

\emph{Notations}: Let $\R$ be the real space, and for any vector $y \in \R^d$, let $\norm{y}$ be its Euclidean norm. For any $\stE \subset \R^m$, let $d(x,\stE) = \inf \{\norm{x-y}: y\in \stE\}$. For any vectors $y_1,\ldots, y_N$ in $\R^d$, the concatenated vector $[y_1^T,\ldots, y_N^T]^T$ is denoted as $\by = \operatorname{Vec}(\{y_i\}_{i=1}^N)$. We also use $\V(A)$ to denote the vector formed by stacking the columns of the matrix $A$ together. We let $\bigangle{\by} \triangleq (y_1+\ldots+y_N)/N \in \R^d$, be the average vector of $y_1,\ldots, y_N$. We denote by $\by_\bot \triangleq \by - \mathbf{1}\otimes \bigangle{\by} $ the disagreement vector, where $\mathbf{1}$ is the vector containing all ones, and $\otimes$ is the Kronecker product. We use $\langle \cdot,\cdot \rangle$ to denote inner product, $\nabla_x f(x)$ to be the gradient with respect to (w.r.t.) $x$ of the function $f$, and $I_d$ to denote the $d\times d$ identity matrix.

\section{System model}
\label{sec:system_model}

\begin{figure}[!htb]
  \centering
  \includegraphics[width=0.35\textwidth]{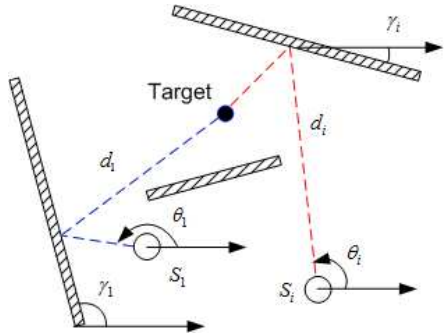}
  \caption{An example for the one-bounce reflection path from the target to sensor $i$ and sensor 1.}
  \label{fig:Figure-Scatter}
\end{figure}

We consider a network of $N$~nodes, where each node has either one single-bounce NLOS signal path or a LOS signal path to the target, as shown in Figure~\ref{fig:Figure-Scatter}. The target could for example be a cell phone user to be localized and tracked, and the sensor nodes diverse picocell base stations deployed in the surrounding area. 
Let $d_i$ be the length of the signal path from the target to node~$i$, $\theta_i$ be the AOA of the signal path at node~$i$, and $\gamma_i$ be the orientation of the scatterer from which the signal path from the target to node~$i$ bounces off. If the signal path is a LOS path, we take the scatterer orientation to be the same as the AOA $\theta_i$. All angles are measured with respect to the horizontal direction. It can be shown \cite{LenTayQue:C12} that 
\begin{align*}
d_i = \bg(\theta_i,\gamma_i)^T(\bq - \bp_i)
\end{align*}
where $\bq$ is the target location, $\bp_i$ denotes the position of the $i$-th node and $\bg(\theta_i,\gamma_i)$ is defined as
\begin{align*}
\bg (\theta_i,\gamma_i)\triangleq \frac{1}{\cos (\theta_i -\gamma_i)}\begin{bmatrix} \cos\gamma_i \\ \sin\gamma_i \end{bmatrix}.
\end{align*}
Without loss of generality, sensor~1 is selected as the reference node, and every other sensor $i\geq 2$ computes a TDOA measurement w.r.t.\ to node~1 given by
\begin{align*}
\td_{i1} = \bg(\theta_i,\gamma_i)^T(\bq-\bp_i) - \bg(\theta_1,\gamma_1)^T(\bq-\bp_1)  + \tn_{i}, 
\end{align*}
where $\tn_{i}$ is the TDOA measurement noise at node~$i$. The measurement noises are assumed to be independent zero-mean Gaussian random variables with $\tn_i$ having variance $\sigma_i^2$ for all $i=1,\ldots,N$. In addition, each node $i\geq 2$ makes a noisy AOA measurement $\tth_i$, modeled by $\tth_i = \theta_i+\eta_i$ where $\eta_i$ is the AOA measurement noise and is independent across nodes. 

\begin{figure}[!htb]
  \centering
  \includegraphics[width=0.45\textwidth]{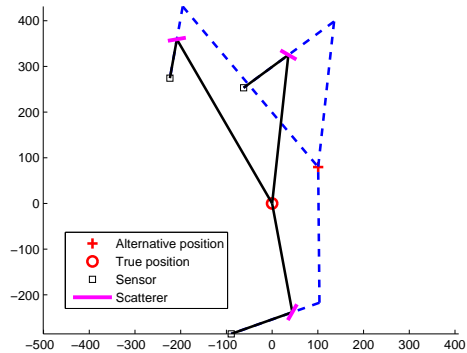}
  \caption{An example in which a set of TDOA and AOA measurements correspond to two possible target locations if no additional information about the scatterers are known. Because any constant added to all the signal path lengths do not change the TDOA values, it is possible to find multiple target locations that satisfy the same AOA measurements.}
  \label{fig:example_multiple_global}
\end{figure}

The orientation angles $\gamma_i$, for $i\geq 2$, are treated as latent variables that are not observed with $\gamma_i \in \Gamma = \{\beta_1, \cdots, \beta_M\}$, where $\beta_j \in [0,2\pi)$ for $j=1,\ldots,M$. However, the reflector orientation of the signal received at sensor 1, is assumed to be known, because otherwise the estimation problem becomes unidentifiable due to the existence of multiple global maximum points in the likelihood function (see Figure \ref{fig:example_multiple_global} for an example). This can be achieved in practice by having sensor 1 estimate the reflector angles in its surroundings using methods in \cite{Lang1999,Creuze2011}.

For the convenience of the reader, we summarize some of the notations we use throughout this paper in Table~\ref{tab:notations}. Several of the notations in Table~\ref{tab:notations} will be defined formally where they first appear in the paper. 

\begin{table}
	\centering
	\caption{Notations used in the paper}
		\begin{tabular}{lc}
			\hline
														\textbf{Symbol} 							& \textbf{Definition} 							\\
			\hline 									
			 											$\theta_i$										& AOA at node $i$ 								\\
			\rowcolor{gray90}			$\gamma_i$ 		  							& Scatterer angle for node $i$ 		\\			
														$d_i$													& Signal path length between target and node $i$	\\
			\rowcolor{gray90} 		$\bq$ 												& location of target 								\\
														$\bp_i$ 											& location of node $i$ 							\\	
			\rowcolor{gray90} 		$\td_{i1}$ 										& TDOA measurement between node $i$ and node $1$ \\
														$\bx_i^n$ 										& parameter vector containing $\bq$ and $\{\theta_i\}^N_{i=1}$ at node $i$ at iteration $n$	\\
			\rowcolor{gray90} 		$z_i$ 								& $z_i= \{\tilde{d}_{i1}, \theta_i, \gamma_i\}$, contains estimates of TDOA, AOA and scatterer angles of node $i$ \\ 
														$\rho_i(\gamma_i,\bx^{n-1})$	& conditional probability of the scatterer angle at node $i$, refer to \eqref{eq:rho_i}			\\
			\rowcolor{gray90}			$S_i(z_i;\theta_1), T_i(z_i;\bq)$	& local statistics at node $i$ (before E-step), refer to \eqref{eq:locLogL_1}-\eqref{eq:locLogL_2}  \\
														$s_i^n,t_i^n$ 								& local statistics at node $i$ (after E-step) at iteration $n$, refer to \eqref{eq:tild_s}-\eqref{eq:s} \\
			\rowcolor{gray90} 		$\lambda_n$ 									& step size at iteration $n$ \\							
														$\phi_1,\phi_2$ 							& local maximizer functions of $\bq$ and $\theta_1$, refer to \eqref{eq:locLogL_1}-\eqref{eq:locLogL_2} \\	
			\rowcolor{gray90} 		$K(\bx)$ 											& log-likelihood function after E-step, refer to \eqref{eq:Kx} \\																								
			\hline
		\end{tabular}
	\label{tab:notations}
\end{table}
\section{EM algorithms for target localization}
\label{sec:em_algo}

In this section, we start by providing a brief review of the standard centralized EM method for target localization, and then describe our distributed EM method. 

\subsection{Centralized EM algorithm}
\label{subsec:centr_em}

\begin{figure}[!htb]
  \centering
  \includegraphics[width=0.25\textwidth]{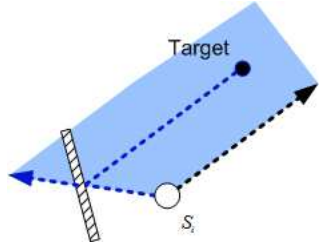}
  \caption{The source lies in the area spanned by the line from node~$i$ in the direction of AOA and the line from node~$i$ in the direction opposite to the AOD. }
  \label{fig:Figure-Scatter-2}
\end{figure}

In the centralized EM method we want to estimate the target location $\bq$, where the reflector orientation $\{\gamma_i\}_{i=2}^N$ is taken to be missing data. Each angle $\gamma_i$ is confined to a support set $\Gamma_i = \{\beta_1,\ldots,\beta_M\}$.  Let $\bx = [\bq^T,\{\theta_i\}_{i=1}^N]^T$ be the parameters of interest. Then, the log likelihood function for the complete data is given by
\begin{align}
&\log\condp{\{\td_{i1}, \tth_i, \gamma_i\}_{i=2}^N,\tth_1}{\bx} \nonumber\\
& = \sum_{i=1}^N\log p(\tth_i|\theta_i) +\sum_{i=2}^N\log \condp{\gamma_i}{\bq,\theta_i,\theta_1}\nonumber\\ 
&+\sum_{i=2}^N \log\condp{\td_{i1}}{\bq, \theta_i,\theta_1,\gamma_i} \label{eq:completeL}
\end{align}
where 
\begin{align*}
&\condp{\td_{i1}}{\bq,\theta_i,\theta_1,\gamma_i} =  \\
&\frac{1}{\sqrt{2\pi\sigma_i^2}} \exp \left(-\frac{1}{2\sigma_i^2}(\td_{i1}+g(\theta_i,\gamma_i)^T\bp_i-g(\theta_1,
\gamma_1)^T\bp_1-(g(\theta_i,\gamma_i)-g(\theta_1,\gamma_1))^T\bq)^2  \right)
\end{align*}
and
\begin{align*}
&\condp{\gamma_i}{\bq,\theta_i,\theta_1} \propto \condp{\bq}{\gamma_i,\theta_i,\theta_1}p(\gamma_i),
\end{align*}
with
\begin{align}
\label{eq:prob_q}
\condp{\bq}{\gamma_i,\theta_i,\theta_1}= 
\begin{cases}
1 & \mbox{ if } \psi_i\in [\theta_i, \phi_i] \\
0 & \mbox{ otherwise } 
\end{cases}
\end{align}
and $\psi_i$ is the angle of $\bq-\bp_i$ w.r.t.\ the horizontal direction. A vector with orientation angle $\phi_i = 2\gamma_i-\theta_i$ w.r.t.\ the horizontal has opposite direction to the vector from the target to the scattering point. The probability \eqref{eq:prob_q} thus restricts the space of possible source locations to the area spanned by the line from node~$i$ in the direction of the AOA and the line from node~$i$ in the direction opposite to the AOD, as shown in Figure~\ref{fig:Figure-Scatter-2}. 

The EM algorithm is an iterative procedure where a new estimate $\bx^n$ is generated at each iteration $n$.  The estimation is done by repeating the following two steps, where $\bx^0$ is an initial guess.
\begin{enumerate}[1)]
\item \emph{E-step}.
The E-step is to account for the missing data, in this case the scatterer orientation angle. At iteration $n$, we have
\begin{align}
&\Qfun{\bx}{\bx^{n-1}}
= \int_{\{\gamma_i\}_{i=2}^N} \log\condp{\{\td_{i1}, \tth_i, \gamma_i\}_{i=2}^N,\tth_1}{\bx} \nonumber\\
&\quad  \cdot \condp{\{\gamma_i\}_{i=2}^N}{\bx^{n-1}, \{\td_{i1}, \tth_i\}_{i=2}^N,\tth_1}\dd \gamma_2\ldots\gamma_N \label{eq:Q1}
\end{align}
where 
\begin{align*}
&\condp{\{\gamma_i\}_{i=2}^N}{\bx^{n-1}, \{\td_{i1}, \tth_i\}_{i=2}^N,\tth_1}\\
&= \prod_{i=2}^N \condp{\gamma_i}{\td_{i1},\bq^{n-1},\theta_i^{n-1},\theta_1^{n-1}}.
\end{align*}
Let us define
\begin{align}
&\rho_i(\gamma_i,\bx^{n-1}) \triangleq
\condp{\gamma_i}{\td_{i1},\bq^{n-1},\theta_i^{n-1},\theta_1^{n-1}} \nonumber\\
&\propto \condp{\td_{i1}}{\bq^{n-1}, \theta_i^{n-1},\theta_1^{n-1},\gamma_i}
\condp{\gamma_i}{\bq^{n-1},\theta_i^{n-1},\theta_1^{n-1}}.
\label{eq:rho_i}
\end{align}
Equation \eqref{eq:Q1} can then be rewritten as 
\begin{align*}
&\Qfun{\bx}{\bx^{n-1}} = \sum_{i=1}^N\log p(\tth_i|\theta_i) \\
&+\sum_{i=2}^N \int_{\gamma_i} \log\condp{\gamma_i}{\bq,\theta_i,\theta_1} \rho_i(\gamma_i,\bx^{n-1})\dd \gamma_i \\
&+\sum_{i=2}^N \int_{\gamma_i} \log\condp{\td_{i1}}{\bq, \theta_i,\theta_1,\gamma_i} \rho_i(\gamma_i,\bx^{n-1})\dd \gamma_i.
\end{align*}

\item \emph{M-step}.
A new estimate of $\bx^n$ is obtained by maximizing the function $Q(\bx, \bx^{n-1})$, i.e.,
\begin{align*}
\bx^n = \arg\max_{\bx} \Qfun{\bx}{\bx^{n-1}}.
\end{align*}
\end{enumerate}

The EM algorithm is not guaranteed to converge to the optimal solution, as it can get stuck in a local maximum. In practice, one can initialize the EM algorithm at different values for $\bx^0$, and if the different runs yield different solutions, select the result with the highest log-likelihood value. 

\subsection{Distributed EM algorithm}
\label{subsec:distr_gem}

We now propose a distributed EM algorithm based on the algorithm described in Section \ref{subsec:centr_em}, and the distributed EM algorithm of \cite{Morral2012, Cappe2009}, which require that the local likelihood functions in \eqref{eq:completeL} belong to the exponential family. We note that our local likelihood functions do not belong to the exponential family of distributions if both the target location $\bq$ and AOA $\theta_1$ are parameters to be estimated. To overcome this difficulty, we write the local likelihood function in two different ways: each time keeping either $\bq$ or $\theta_1$ (but not both) as the single parameter of interest. By performing this decomposition, and a sequential optimization in the M-step, it turns out that the procedure of \cite{Morral2012, Cappe2009} can be adapted to perform distributed estimation of our target location. However, the convergence of our proposed method does not follow directly from the analysis in \cite{Morral2012, Cappe2009}, and we require further technical assumptions and work in our convergence analysis, although similar conclusions as in \cite{Morral2012, Cappe2009} are derived.

We assume that each sensor $i$ knows its own location $\bp_i$ and sensor 1 broadcasts its information $(\bp_1, \gamma_1, \tth_1)$ to all sensors. We further assume that the AOA noise $\eta_i$ follows a uniform distribution, i.e., Unif$[-\eta^0,\eta^0]$. Under these assumptions, the first two terms in the complete data likelihood given by \eqref{eq:completeL} can be converted to constraints on $\{\theta_i\}_{i=1}^N$ and $\bq$. 

Since the TDOA measurement noise is Gaussian, the local log-likelihood function of node $i$ is given by 
\begin{align*}
&\log\condp{\td_{i1}}{\bq, \theta_1,\theta_i,\gamma_i} \\
&= -\frac{1}{2\sigma_i^2}\left(\td_{i1} - \bg(\theta_i,\gamma_i)^T(\bq-\bp_i) + \bg(\theta_1,\gamma_1)^T (\bq-\bp_1)\right) ^2 \\
\end{align*}
After some algebraic manipulations to isolate terms that depend on $\bq$ or $\theta_1$ only and defining $z_i= \{\tilde{d}_{i1}, \theta_i, \gamma_i\}$, the local log-likelihood can be rewritten as follows: 
\begin{align}
&\log\condp{\td_{i1}}{\bq, \theta_1,\theta_i,\gamma_i} \nonumber\\
& = c_{1,i}(z_i,\theta_1) +  S_i(z_i;\theta_1)^T \phi_{1}(\bq) \label{eq:locLogL_1}\\
& = c_{2,i}(z_i,\bq) + T_i(z_i;\bq)^T \phi_{2}(\theta_1) \label{eq:locLogL_2}\\
& = \psi_i(z_i, \theta_i, \bq, \theta_1) \label{eq:locLogL_3}
\end{align}
where $c_{1,i}(\cdot)$ and $c_{2,i}(\cdot)$ are normalizing factors, and $\phi_1(\bq) = [\V(\bq\bq^T);\bq]$, $\phi_2(\theta_1) = [1/\cos^2(\theta_1-\gamma_1),1/\cos(\theta_1-\gamma_1)]^T$, and
\begin{align*}
&S_i(z_i;\theta_1)= 
\begin{bmatrix}
\V(\bU_i) \\
-2\bV_i
\end{bmatrix}
\end{align*}
with
\begin{align*}
\bU_i & = -\frac{1}{2\sigma_i^2}(\bg(\theta_i,\gamma_i)-\bg(\theta_1,\gamma_1))(\bg(\theta_i,\gamma_i)-\bg(\theta_1,\gamma_1))^T, \\
\bV_i & = -\frac{1}{2\sigma_i^2} (\td_{i1}+\bg(\theta_i,\gamma_i)^T \bp_i - \bg(\theta_1,\gamma_1)^T\bp_1)\\
&\hspace{.4cm}\cdot(\bg(\theta_i,\gamma_i)-\bg(\theta_1,\gamma_1)),
\end{align*}
and 
\begin{align*}
&T_i(z_i;\bq) = -\frac{1}{2\sigma_i^2} \\
&\cdot \begin{bmatrix}
\left(\begin{bmatrix}\cos\gamma_1\\ \sin\gamma_1\end{bmatrix}^T(\bq-\bp_1)\right)^2 \\
2(\td_{i1}-\bg(\theta_i,\gamma_i)^T(\bq-\bp_i))\begin{bmatrix}\cos\gamma_1\\ \sin\gamma_1\end{bmatrix}^T(\bq-\bp_1)
\end{bmatrix}.
\end{align*}

\medskip

The fundamental idea of the distributed EM algorithm is to adapt the local log-likelihood functions \eqref{eq:locLogL_1}-\eqref{eq:locLogL_3} of the different nodes so that they approximate the complete log-likelihood function in \eqref{eq:completeL}. During each iteration of the distributed EM algorithm, the nodes perform the following three steps: the nodes first perform a local E-step to account for the missing scatterer orientation data. During the gossip step, the nodes exchange appropriate statistics so that their local log-likelihood functions approximate the complete log-likelihood function. Finally, each node performs a local M-step to determine the parameters $\bx = [\bq^T,\{\theta_i\}_{i=1}^N]^T$ that maximize its log-likelihood function. Under the technical conditions given in Assumptions \ref{assum:pdf}, \ref{assum:lambda} and \ref{assum:wn} below, the estimates of all the nodes converge to an identical solution. In the following, we let $\bx_i^n = [(\bq_i^n)^T, \theta_{1,i}^n, \theta_i^n]^T$ be the local estimate at node $i$ of the parameters $[\bq^T,\theta_1,\theta_i]$ at iteration $n$. The distributed EM algorithm is given as follows:

\begin{enumerate}

 \item \textit{Local E-step: } At each iteration, each node accounts for its missing data (i.e., the orientation angle of the scatterer from which its received signal bounced off): 

\begin{align*}
\bars_i(\bx)  &= \int_{\gamma_i} S_i(z_i;\theta_1) \rho_i(\gamma_i,\bx)\dd \gamma_i, \\ 
\bart_i(\bx) 	&= \int_{\gamma_i} T_i(z_i;\bq) \rho_i(\gamma_i,\bx)\dd \gamma_i, \\
\bar{\psi}_i(\theta_i,\bq,\theta_1;\bx) &= \int_{\gamma_i} \psi_i(z_i,\bq,\theta_1) \rho_i(\gamma_i,\bx)\dd \gamma_i.
\end{align*}

Each node $i$ then the local statistics $\tilde{s}_i^n$ and $\tilde{t}_i^n$: 
\begin{subequations}
\begin{align}
\ts_i^n & = s_i^{n-1} + \lambda_{n}[\bars_i(\bx_i^{n-1})-s_i^{n-1}], \\
\tilt_i^n & = t_i^{n-1} + \lambda_{n}[\bart_i(\bx_i^{n-1})-t_i^{n-1}],
\end{align}\label{eq:tild_s}
\end{subequations}
where $\lambda_n \geq 0$ is a chosen step size. This update process is to ensure that information from the other nodes obtained in previous iterations (contained in $s_i^{n-1}$ and $t_i^{n-1}$) is maintained during the current iteration. 
 
 \item \textit{Gossip step: } Each node $i$ broadcasts $\{\ts_i^n,\tilde{t}_i^n\}$ to its neighbors, and collects $\{\ts_j^n,\tilde{t}_j^n\}$ from its neighboring sensors $j$. It then computes the weighted average: 
\begin{subequations}
\begin{align}
s_i^n &= \sum_{j=2}^N w_n(i,j)\ts_j^n \\
t_i^n &= \sum_{j=2}^N w_n(i,j)\tilt_j^n
\end{align} \label{eq:s}%
\end{subequations} 
where $w_n(i,j)$ are non-negative weights, with $w_n(i,j)=0$ if node $j$ is not a neighbor of node $i$. This step enables node $i$ to approximate the the log-likelihood function in the last term in \eqref{eq:completeL} (note that the first two terms in \eqref{eq:completeL} have been converted to constraints on $\{\theta_i\}_{i=1}^N$ and $\bq$). 
%If more nodes communicate with each other during the gossip step, the local log-likelihood approximations will converge faster to the complete log-likelihood function in \eqref{eq:completeL}. 
  
 \item \textit{Local M-step: } Each node determines the value of $\bx_i^n$ that maximizes its local log-likelihood function as follows: 
\begin{align}
\bq_i^n & = \max_{\bq\in A_i}\ (s_i^n)^T \phi_1(\bq), \label{max_q}\\
\theta_{1,i}^n & = \max_{\theta_1\in \Theta_1} \  (t_i^n)^T \phi_2(\theta_1), \label{max_theta1}\\
\theta_i^n & = \max_{\theta_i\in\Theta_i}\bar{\psi}_i(\theta_i,\bq_i^n,\theta_{1,i}^n; \bx_i^{n-1}),\label{max_thetai}
\end{align} 
where $A_i = \{\bq : p(\bq | \gamma, \theta_i, \theta_1) > 0\text{ for }\gamma \in \Gamma\}$ and $\Theta_i= [\tth_i-\eta^0, \tth_i+\eta^0]$ are the constraint sets corresponding to the first two terms of \eqref{eq:completeL}. 
Optimizing separately over each parameter $\bq$, $\theta_1$ and $\theta_i$ is much easier than performing a joint maximization over all three parameters, as each optimization in \eqref{max_q}-\eqref{max_thetai} has a unique global maximum. We will see in Theorem \ref{th:conv_max} that this also ensures that the estimated $\bq$ at every node converges to the same value. 

\end{enumerate}
In the following section, we investigate the assumptions under which the proposed distributed EM algorithm converges.

\section{Convergence analysis}
\label{sec:conv_anal}

We require the following technical assumptions to show that the proposed distributed EM algorithm converges. These assumptions are similar to those in \cite{Morral2012, Cappe2009}, which however requires that \eqref{eq:locLogL_3} belongs to the exponential family.
\begin{assumption}\label{assum:pdf}
There are convex open subsets $\stS$ and  $\stT$ such that:
\begin{enumerate}[(a)]
\item  
for any $s_i\in \stS$, and $\lambda\in[0,1]$, $(1-\lambda)s_i+\lambda\bars_i(\bx))\in\stS$;
for any $t_i\in \stT$, and $\lambda\in[0,1]$, $(1-\lambda)t_i+\lambda\bart_i(\bx))\in\stT$.

\item
for any $s_i\in \stS$, the function $\bq\mapsto s_i^T \phi_{1}(\bq) $ has a unique global maximum denoted by $\bar{\bq}(s_i)$;
for any $t_i\in \stT$, the function $\theta_1\mapsto t_i^T\phi_{2}(\theta_1)$ has a unique global maximum denoted by $\barth_1(t_i)$;
for any $s_i\in \stS$, any $t_i\in \stT$, the function $\theta_i \mapsto \bar{\psi}_i(\theta_i,\bar{\bq}(s_i),\barth_1(t_i);\bx^{n-1})$ has a unique global maximum.
\end{enumerate}
\end{assumption}
Note that Assumption \ref{assum:pdf}(b) may not hold when $\bar{\bq}(s)$ or $\bar{\theta}_1(t)$ hit the boundaries of its domain. In addition, Assumption \ref{assum:pdf}(a) is hard to verify in practice. Therefore, it is of common practice to confine $\ts_i$ and $\tilt_i$ in the updating step \eqref{eq:tild_s} to the given convex sets $\stS$ and $\stT$ by projection. The projection procedure does not affect the convergence analysis.

\begin{assumption}\label{assum:lambda}
The step sizes $\lambda_n$ are chosen so that
$\sum_{n=1}^\infty \lambda_n = \infty$ and $\sum_{n=1}^\infty \lambda_n^2 < \infty$. 
\end{assumption}

\begin{assumption} \label{assum:wn}
The weighting matrix has the following properties.
\begin{enumerate}[(a)]
\item For any $n$, $W_n \triangleq [w_n(i,j)]_{i,j = 2\ldots N} $ is a matrix-valued random variable with non negative elements.
\item For any $n$, $W_n$ is row stochastic : $W_n \mathbf{1} = \mathbf{1}$.
\item $W_n$ is column stochastic in expectation: $\E(W_n)^T\mathbf{1} = \mathbf{1}$.
\item $\{W_n\}_{n\geq 1}$ is an independent identically distributed sequence.
\item The spectral norm $\rho$ of matrix $\E(W_n^T(I_N-\mathbf{1}\mathbf{1}^T/N)W_n)$ satisfies $\rho<1$
\end{enumerate}
\end{assumption}

Finally, let us define $K(\bx)$ as
\begin{align}
K(\bx) &\triangleq \log\int\condp{\{\td_{i1}, \tth_i, \gamma_i\}_{i=2}^N,\tth_1}{\bx} \dd \gamma_2\ldots, \gamma_N
\label{eq:Kx}
\end{align}

We let $\stE = \{\bx: \nabla _\bx K(\bx) = 0\}$. The following result shows that the sequence $\bx_i^n$ in our distributed EM algorithm converges to the same local maximizer for all sensors. 

\begin{theorem} 
\label{th:conv_max}
Under Assumptions \ref{assum:pdf}, \ref{assum:lambda} and \ref{assum:wn}, we have $\lim_{n\to \infty}\max_{i,j = 2\ldots N} \norm{\bx_i^n-\bx_j^n} = 0$ with probability 1 and  $\lim_{n\to \infty}d(\bx_i^n,\stE)=0$ with probability 1 for $i=2,\ldots,N$. 
\end{theorem}

We provide an outline for the proof of Theorem~\ref{th:conv_max} below, largely based on the proof of Theorem~1 in \cite{Bianchi2013}. The full proof is provided in Appendix~\ref{sec:proof_theorem}.

\begin{enumerate}[(i)]
	\item We start by showing that the nodes asymptotically reach a consensus on their estimate, i.e., $\lim_{n\to \infty}\max_{i,j = 2\ldots N} \norm{\bx_i^n-\bx_j^n} = 0$ with probability 1. Therefore, the convergence analysis of the vectors $\bs_n=\operatorname{Vec}(\{s_i^n\}_{i=2}^N)$ and $\bt_n =  \operatorname{Vec}(\{t_i^n\}_{i=2}^N)$ reduces to analyzing the average estimates $\bigangle{\bs_n}$ and $\bigangle{\bt_n}$ (see Lemma 1 in \cite{Bianchi2013}). 
	\item We show that $\bigangle{\bs_n}$ and $\bigangle{\bt_n}$ follow the following time-dynamical system: 
\begin{align*}
&\bigangle{\bs_n} = \bigangle{\bs_{n-1}}+\lambda_n h_1(\bigangle{\bs_{n-1}},\bigangle{\bt_{n-1}})+\lambda_n \xi_1^n +\lambda_n r_1^n \\
&\bigangle{\bt_n} = \bigangle{\bt_{n-1}}+\lambda_n h_2(\bigangle{\bs_{n-1}},\bigangle{\bt_{n-1}})+\lambda_n \xi_2^n +\lambda_n r_2^n 
\end{align*}
where $h_1(\cdot,\cdot)$ and $h_2(\cdot,\cdot)$ are two functions defined in \eqref{eq:h_12_st}, while $\xi_l^n$ and $r_l^n$ for $l=1,2$ are error terms given by \eqref{eq:xi_r}. These equations can be viewed as noisy approximations of the following Ordinary Differential Equation (ODE): 
\begin{align*}
&\dot{s} = h_1(s,t)\\
&\dot{t}= h_2(s,t)
\end{align*}
We show that the roots of $h_1$ and $h_2$ correspond to a stationary point of $\bx\mapsto K(\bx)$.
	\item We then show that the sequence of $\bx_i^n$ converges to the same local maximizers for all nodes. 
	\item Finally, we show that the terms $\lambda_n \xi_1^n$, $\lambda_n \xi_2^n$, $\lambda_n r_1^n$ and $\lambda_n r_2^n$ asymptotically tend to 0 with probability 1. 
\end{enumerate}
By combining these results, we show that our distributed EM algorithm converges to the same local maximizer for all nodes. 
\section{Simulation results}
\label{sec:simul}

In this section, we present simulation results to verify the performance of our proposed distributed EM algorithm. In our simulations, we use the pairwise gossip scheme \cite{Boyd2006} in the Gossip Step of the distributed EM algorithm. Specifically, at each iteration, two neighboring nodes $i$ and $j$ are randomly chosen to compute the weighted averages $s_i^n = s_j^n =0.5\tilde{s}_i^n+ 0.5\tilde{s}_j^n$ and $t_i^n = t_j^n =0.5\tilde{t}_i^n+ 0.5\tilde{t}_j^n$. For other nodes $k\not\in\{i,j\}$, $s_k^n = \tilde{s}_k^n$ and $t_k^n = \tilde{t}_k^n$. 

We compare the performance of the proposed algorithm with the centralized EM algorithm, and an algorithm using TDOA only in order to show the impact of additional AOA information on the localization accuracy. The \emph{TDOA only} target estimation is formulated as 
\begin{align}\label{eq:tdoa}
\min_{\{d_i\}_{i=1}^N,\bq} \hspace{4pt} \sum_{i=2}^N(\frac{d_i-d_1-\td_{i1}}{\sigma_i})^2+\delta d_1^2\nonumber\\
s.t.\quad \|\bq-\bp_i\|\leq d_i,\quad i= 1,\ldots,N
\end{align}
where $\delta$ is a chosen positive constant.

\begin{figure}[!ht]
  \centering
  \includegraphics[width=0.45\textwidth]{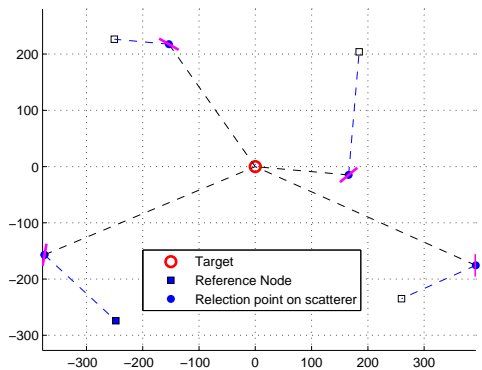}
  \caption{Configuration of sensors and  the target. The dotted lines denote the signal paths between the target and sensors.}
  \label{fig:scenario}
\end{figure}
\begin{figure}[!ht]
  \centering
  \includegraphics[width=0.45\textwidth]{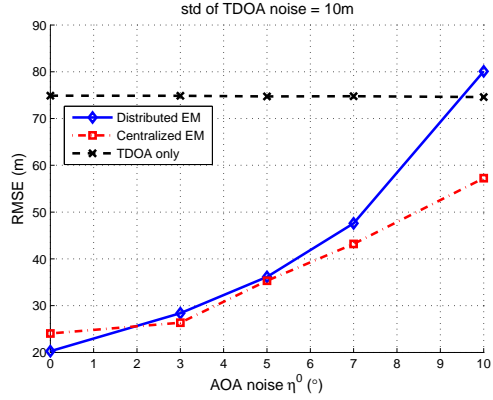}
  \caption{RMSE of the estimated target position versus angle noise $\eta^0$, with $\sigma_i = 10$m, $i=1,\ldots,N$.}
  \label{fig:AOARMSE}
\end{figure}
\begin{figure}[!ht]
  \centering
  \includegraphics[width=0.45\textwidth]{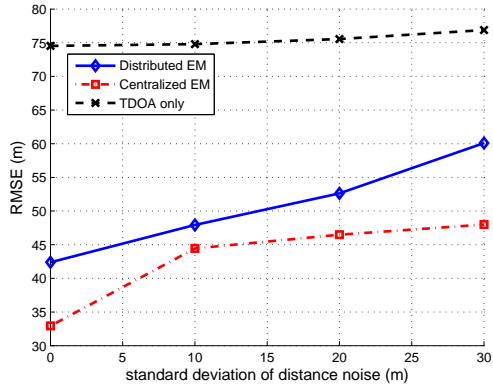}
  \caption{RMSE of the estimated target position versus standard deviation of distance noise $\sigma_i$, with $\eta^0 = 7^\circ$.}
  \label{fig:distRMSE}
\end{figure}

The simulated scenario is shown in \figurename~\ref{fig:scenario}. The square markers represent the sensors. Taking the support set of $\gamma_i$ in our algorithm to be $[\gamma_i^0-10^\circ,\gamma_i^0+10^\circ]$, where $\gamma_i^0$ is the true scatterer orientation, and the angle noise $\eta_i$ uniformly distributed in $[-\eta^0, \eta^0]$, the root mean square error (RMSE) of the estimated target position is shown in \figurename~\ref{fig:AOARMSE} and \figurename~\ref{fig:distRMSE}. Comparing the performances of our distributed EM and \emph{TDOA only} methods, it is observed that the localization accuracy is greatly improved by using the additional AOA information. It is also seen that our proposed algorithm has similar performance as that of the centralized EM method, except when the measurement noises become large.

\section{Experimental results}
\label{sec:exper}

In this section, we evaluate the performance of the distributed EM algorithm in an experimental setting. We start by describing the measurement setup, and then analyze the localization performance. Note that the TDOA and AOA measurements are separate experiments, performed with the nodes placed in identical locations. The TDOA and AOA at each node are extracted from the measurements for each transmitter location, and fused off-line to evaluate the performance of the distributed EM algorithm. 

\subsection{Measurement setup}
\label{subsec:meas_setup}

\begin{figure}[!ht]
  \centering
  \includegraphics[width=0.35\textwidth]{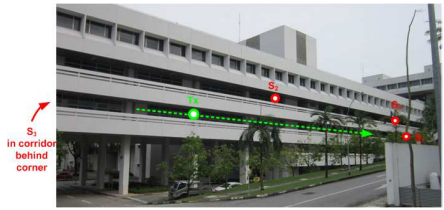}
  \caption{Picture of the measurement site. The two NLOS nodes are placed on a different floor than the transmitter and the LOS node. The node $S_3$ is around a corner of the corridor in which the transmitter is placed. }
  \label{fig:picture_meas}
\end{figure}

We consider a network of 1~transmitting target and 4~sensor nodes, located on different floors of a building facing another building about 40 m away. The transmitter and one of the sensors ($S_1$) are placed on a corridor on the same floor with a clear LOS. Another sensor ($S_3$) is placed on the same floor as the transmitter, but behind a corner in a NLOS situation. The signal path from the transmitter to $S_3$ is expected to be a diffracted path around the corner. Although this signal path does not have a single reflection from a scatterer, as assumed in our system model, the diffracted path can nevertheless be modeled as a reflected path off a virtual scatterer near the corner. Two of the sensors, $S_2$ and $S_4$, are placed on a different floor so that the signal path from the target to each of these nodes consist of a NLOS path bounced off the opposite building (see Figure~\ref{fig:picture_meas}). The transmitter was moved along 13~different measurement locations along a line going from $S_1$ towards $S_3$. 
The transmitter is moved to 13~different measurement locations along a line, as shown in Figure~\ref{fig:picture_meas}.  
\begin{figure}[!ht]
  \centering
  \includegraphics[width=0.4\textwidth]{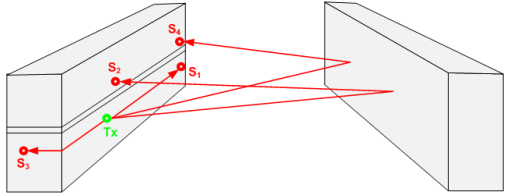}
  \caption{Expected signal paths between the transmitter and the different receiver nodes. }
  \label{fig:drawing_meas}
\end{figure}
The distance between node $S_1$ and and the corridor leading to $S_3$ is 100~m, and the distance between nodes $S_2$ and $S_4$ is 50~m. The longest path between any transmitter location and any receiver is 106~m. The largest measured TDOA is 160~m. The building opposite the measurement building is approximately 40~m away. 

{\bf AOA measurement: }A 4-channel Pentek 4995A A/D converter is used as a 4-element MIMO receiver. The A/D converter is able to sample signals up to 200~MHz, and is used as a sub-sampled receiver by connecting the antennas straight to the A/D converter~\cite{Kim2010}. The source signal is generated with a signal generator sending a pilot tone at 795~MHz. The four antennas of the receiver form a linear array with a distance of 15~cm between antennas. We use the well-known Multiple Signal Classification (MUSIC) algorithm to recover the AOA of the signal paths~\cite{Schmidt1986}. For a 4-element array, the MUSIC algorithm is able to recover up to three signal paths, even though in our experiment, there is only be one dominant path from the target to each sensor node. 
The receiver has been carefully calibrated, and measurements in an anechoic chamber show that our setup has an zero-mean AOA error with a standard deviation of $3^{\circ}$. 
%For each receiver node location, the transmitter is placed in all 13~measurement locations, and the AOA at the receiver is measured. 
Our AOA measurements confirm that for nodes $S_2$ and $S_4$, the dominant path from the transmitter to the NLOS nodes occurs through a reflection on the building opposite the receiver nodes. For node $S_3$, the measurement confirm that the dominant path comes through diffraction around the corner of the building. 

{\bf TDOA measurement: }We use a USRP software-defined radio platform to measure the TDOA between different nodes. The USRPs are equipped with WBX daughterboards and a GPSDO module, which synchronizes the internal clock and local oscillator of the different nodes to the GPS UTC time~\cite{usrp2013}. However, even with GPS synchronization, the time offset between the different nodes can still be as large as 100~ns, causing TDOA errors up to 60~m. To improve on this, we use a simple relaying architecture described in Appendix~\ref{sec:tdoa_meas}. We characterize our TDOA measurement setup by measuring the TDOA between different nodes in an ideal environment (with cables, no multipath and high SNR) and find that the TDOA measurement error is as small as 3~m. In outdoor line-of-sight environments, the TDOA error of our setup has a mean of 1~m and a standard deviation of 4~m. For each transmitter location, 10~TDOA measurements are taken. Similarly as for the AOA measurement, the measured values for our TDOA indicate that for the NLOS nodes the dominant path occurs through a reflection on the building opposite the receiver nodes or through diffraction around the corner of the building. All the parameters of our setup are given in Appendix~\ref{sec:tdoa_meas}. We stress that our TDOA measurement setup \textit{does not} aim at reproducing the distributed EM algorithm, but is used to measure the TDOA between different pair of nodes so that the distributed EM algorithm can be evaluated over real measurements offline.

\subsection{Localization results}
\label{subsec:local_results}

\begin{figure}[!ht]
  \centering
  \includegraphics[width=0.45\textwidth]{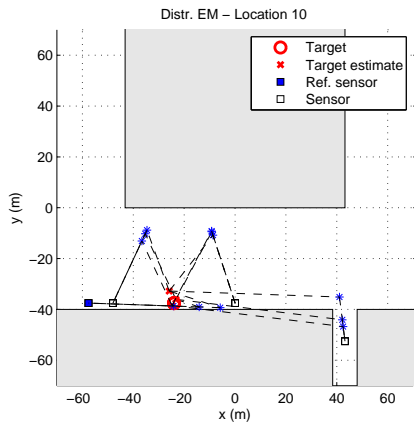}
  \caption{Localization result when using the distributed EM algorithm. The scatterer angle support set was chosen to be $[\gamma_i^0-10^\circ,\gamma_i^0+10^\circ]$. }
  \label{fig:distrEM_smallAngleSupportSet_loc10}
\end{figure}

For each of the 13 target locations, we make ten TDOA measurements and one AOA measurement. We use the distributed EM algorithm to determine the location of the target for each of these $13\times 10$~measurements. We also compute the location estimate with the centralized EM algorithm to serve as a benchmark. Figure~\ref{fig:distrEM_smallAngleSupportSet_loc10} shows an example of the algorithm results for one particular transmitter location. In this case, the scatterer angle support set was chosen to be $[\gamma_i^0-10^\circ,\gamma_i^0+10^\circ]$, where $\gamma_i^0$ is the true scatterer orientation. Even though the algorithm does not know the building location, it is able to obtain a fair estimate of the scatterer locations, and to estimate the target location. 
It can be observed how the LOS node $S_1$ is treated exactly as a NLOS node. In that case, the algorithm chooses a scatterer whose orientation is parallel to the line between transmitter and LOS node. 

\medskip

\begin{figure}[!ht]
  \centering
  \includegraphics[width=0.45\textwidth]{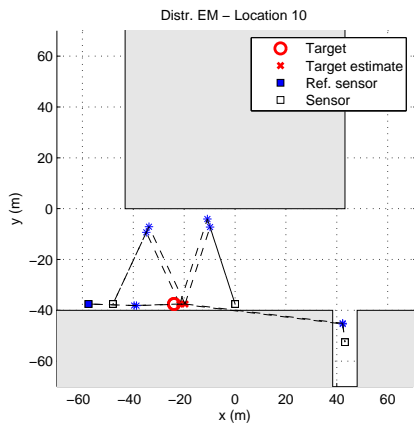}
  \caption{Localization result when using the distributed EM algorithm. The scatterer angle support set was chosen to be $[0^\circ,90^\circ,135^\circ,180^\circ,270^\circ]$. }
  \label{fig:distrEM_uniqueAngleSupportSet_loc10}
\end{figure}

Assuming knowledge of the scatterer orientation angle may be an unrealistic assumption in many scenarios. Figure~\ref{fig:distrEM_uniqueAngleSupportSet_loc10} shows an example of the algorithm output when the scatterer angle support set was chosen to be $[0^\circ,90^\circ,135^\circ,180^\circ,270^\circ]$. In this scenario, the possible scatterer orientation angles are limited to a finite set, which is a fairly realistic assumption in environments with symmetric geometries (e.g. cities with regular city blocks). It can be seen that this scenario the distributed EM algorithm is able to successfully locate the RF target. 

\medskip

\begin{figure}[!ht]
  \centering
  \includegraphics[width=0.45\textwidth]{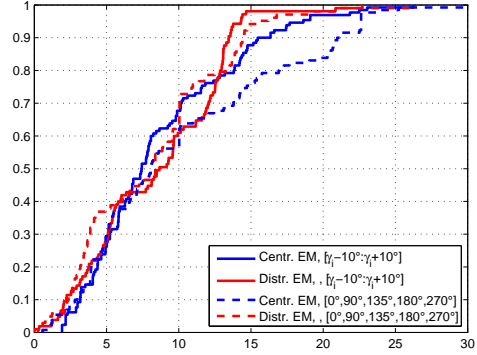}
  \caption{Cumulative distribution function of the localization error, both for the centralized and distributed EM algorithm for two types of scatterer angle support set. }
  \label{fig:error_EM}
\end{figure}

Figure~\ref{fig:error_EM} shows the localization error over all measurements, both for the centralized and distributed EM (the results for the two types of scatterer angle support sets are shown). Note that for the distributed EM, the algorithm could converge in 82\% of the measurements, while for the centralized EM, the algorithm could always converge. The cases where the distributed EM could not converge correspond to cases where the TDOA and/or AOA measurements are very bad, and the estimates of the different nodes diverge too strontly for the algorithm to converge to a unique solution. In the centralized case, these measurements just result in poor localization accuracy. In Figure~\ref{fig:error_EM}, the measurements with large errors in the centralized EM algorithm correspond to the measurements for which our distributed EM algorithm could not find a solution, which explaines why the distributed EM slightly outperforms the centralized EM. For all measurements, the localization error is below 15~m. 

\section{Conclusion}
\label{sec:conclu}

In this work, we propose a distributed EM algorithm for target localization in NLOS environments. We provide sufficient conditions for the proposed algorithm to converge, and prove its convergence. Simulation results show that additional AOA information which is utilized in the proposed EM algorithm significantly improves localization accuracy. When applying the distributed EM algorithm to experimental measurements, it is observed that the algorithm is able to localize a target with an accuracy below 15~m, and has errors similar to a centralized EM approach, verifying that our algorithm works when applied to realistic environments. 

\appendices
\section{Proof of Theorem \ref{th:conv_max}}
\label{sec:proof_theorem}

In this section we provide a proof for Theorem~\ref{th:conv_max}, based on the outline detailed in Section~\ref{sec:conv_anal}. In the sequel, we will use following notations. We let $\bs_n =  \operatorname{Vec}(\{s_i^n\}_{i=2}^N)$, $\bt_n =  \operatorname{Vec}(\{t_i^n\}_{i=2}^N)$, $\bA_n = \operatorname{Vec}(\{\bars_i(\bx_i^{n-1})-s_i^{n-1}\}_{i=2}^N)$ and 
$\bB_n = \operatorname{Vec}(\{\bart_i(\bx_i^{n-1})-t_i^{n-1}\}_{i=2}^N)$. Then the local step \eqref{eq:tild_s} and gossip step \eqref{eq:s} can be combined as 
\begin{subequations}
\begin{align}
&\bs_n = (W_n\otimes I_{d_s}) [\bs_{n-1}+\lambda_n \bA_n]\\
&\bt_n =  (W_n\otimes I_{d_t}) [\bt_{n-1}+\lambda_n \bB_n]
\end{align}
\end{subequations}

The following lemma follows directly from Lemma 1 of \cite{Bianchi2013}. It shows that nodes asymptotically reach a consensus on their estimate. Therefore, the convergence analysis of the vector $\bs_n$ and $\bt_n$ reduce to an analysis of the average estimates $\bigangle{\bs_n}$ and $\bigangle{\bt_n}$. 

\begin{lemma}\label{lem:EMconsensus}
Under Assumptions \ref{assum:pdf}, \ref{assum:lambda} and \ref{assum:wn}, we have $\max_{i,j = 2\ldots N}\norm{s_i^n-s_j^n}$ and \\
$\max_{i,j = 2\ldots N}\norm{t_i^n-t_j^n}$ converging to zero a.s.\ as $n$ tends to infinity.
\end{lemma}

\subsection{Discrete-time dynamical system of $\bigangle{\bs_n}$ and $\bigangle{\bt_n}$}

We now proceed to the analysis of the average estimates $\bigangle{\bs_n}$ and $\bigangle{\bt_n}$. The terms $\bigangle{\bs_n}$ and $\bigangle{\bt_n}$ follow the below discrete time dynamical system:
\begin{subequations}
\label{eq:st_sp_1}
\begin{align}
&\bigangle{\bs_n} = \bigangle{\bs_{n-1}}+\lambda_n h_1(\bigangle{\bs_{n-1}},\bigangle{\bt_{n-1}})+\lambda_n \xi_1^n +\lambda_n r_1^n \\
&\bigangle{\bt_n} = \bigangle{\bt_{n-1}}+\lambda_n h_2(\bigangle{\bs_{n-1}},\bigangle{\bt_{n-1}})+\lambda_n \xi_2^n +\lambda_n r_2^n 
\end{align}
\end{subequations}
with
\begin{subequations}
\label{eq:h_12_st}
\begin{align}
&h_1(s,t) \triangleq \bars(\bar{\bq}(s),\barth_1(t))-s\\
&h_2(s,t) \triangleq \bart(\bar{\bq}(s),\barth_1(t))-t 
\end{align}
\end{subequations}
\begin{subequations}
\begin{align}
&\bars(\bar{\bq}(s),\barth_1(t)) \triangleq \frac{1}{N-1}\sum_{i=2}^N \bars_i(\bar{\bq}(s),\barth_1(t))\\
&\bart(\bar{\bq}(s),\barth_1(t)) \triangleq \frac{1}{N-1}\sum_{i=2}^N \bart_i(\bar{\bq}(s),\barth_1(t))
\end{align}\label{eq:st_bar}
\end{subequations}
\begin{subequations}
\begin{align}
&\xi_1^n =\frac{1}{(N-1)\lambda_n} [(\mathbf{1}^T W_n-\mathbf{1}^T)\otimes I_{d_s}] [\bs_{n-1}+\lambda_n \bA_n]\\
&r_1^n = \frac{1}{N-1} \sum_{i=2}^N (\bars_i(\bar{\bq}(s_i^n),\barth_1(t_i^n))-s_i^n) -h_1(\bigangle{\bs_{n-1}},\bigangle{\bt_{n-1}})\\
&\xi_2^n =\frac{1}{(N-1)\lambda_n} [(\mathbf{1}^T W_n-\mathbf{1}^T)\otimes I_{d_t}] [\bt_{n-1}+\lambda_n \bB_n]\\
&r_2^n = \frac{1}{N-1} \sum_{i=1}^N (\bart_i(\bar{\bq}(s_i^n),\barth_1(t_i^n))-t_i^n)-  h_2(\bigangle{\bs_{n-1}},\bigangle{\bt_{n-1}})
\end{align}\label{eq:xi_r}
\end{subequations}

The discrete time dynamical system above can be viewed as a noisy approximation of the following ordinary differential equation (ODE),
\begin{subequations}
\label{eq:ode}
\begin{align}
&\dot{s} = h_1(s,t)\\
&\dot{t}= h_2(s,t)
\end{align}
\end{subequations}
where $\dot{s}$ and $\dot{t}$ respectively denote the derivative of $s$ and $t$ with respect to time. The following proposition shows that 
\begin{align*}
w(s,t) \triangleq -K(\bar{\bq}(s),\barth_1(t))
\end{align*}
is a Lyapunov function of the ODE \eqref{eq:ode}. In other works, Proposition~\ref{prop:1} shows that the roots of $h_1$ and $h_2$ correspond to a stationary point of $\bx\mapsto K(\bx)$, i.e., $\nabla_\bx K(\bx)|_{\bx =\bx^*}=0$;

\begin{proposition}\label{prop:1}
Under Assumptions \ref{assum:pdf} and \ref{assum:lambda}, if $(s^*, t^*)$ is such that $h_1(s^*,t^*) =0$ and $h_2(s^*,t^*) = 0$, then $\bx^* =  [\bar{\bq}(s^*); \barth_1(t^*)]$ is a stationary point of $\bx\mapsto K(\bx)$, i.e., $\nabla_\bx K(\bx)|_{\bx =\bx^*}=0$. Conversely, for any $\bx^*=[\bq^*;\theta_1^*]$, such that $\nabla_\bx K(\bx)|_{\bx= \bx^*}=0$, we have $s^* = \bars(\bx^*)$, 
$t^* = \bart(\bx^*) $ satisfy $h_1(s^*,t^*) =0$ and $h_2(s^*,t^*) = 0$.
\end{proposition}

\begin{IEEEproof}
To simplify notations, let $\bx^*=[\bq^*;\theta_1^*]$ with $\bq^* = \bar{\bq}(s^*)$ and $\theta_1^* = \barth_1(t^*)$. Since the function $\bq \mapsto \phi_{1}^T(\bq)s$ has a unique global maximum at $\bar{\bq}(s)$ and $\theta_1 \mapsto \phi_2 ^T (\theta_1) t$ has a unique global maximum at $\barth_1(t)$ (cf.\ Assumption \ref{assum:pdf}(c)), we obtain
\begin{align}\label{eq:uniStation}
\nabla_{\bq} \phi_1^T(\bq^*)s^* =0, 
\quad \nabla_{\theta_1} \phi_2 ^T (\theta_1^*) t^* = 0.
\end{align}
From Fisher's identity \cite{Dempster1977}, we have
\begin{align}\label{eq:derivativeK}
\nabla_\bx K(\bx) &= \int\left\{\nabla_\bx\log\condp{\{\td_{i1},\theta_i,\gamma_i)\}_{i=2}^N,\tth_i}{\bx}\right\}\dd \gamma_2\ldots,\gamma_N\dd \theta_2\ldots,\theta_N\nonumber\\
&= (N-1)\begin{bmatrix}
\nabla_{\bq} \phi_1^T(\bq)\bars(\bx)
\\
\nabla_{\theta_1} \phi_2^T(\theta_1)\bart(\bx)
\end{bmatrix}
\end{align}
where the $N-1$ factor arises because $\bars$ and $\bart$ defined in \eqref{eq:st_bar} are normalized by $N-1$. Without loss of generality, we will omit the factor $N-1$ for conciseness. 

Since $h_1(s^*,t^*)=0$ and $h_2(s^*,t^*)=0$, i.e., $s^* = \bars(\bx^*)$, 
$t^* = \bart(\bx^*) $, \eqref{eq:uniStation} and \eqref{eq:derivativeK} imply that
\begin{align*}
\nabla_\bx K(\bx)|_{\bx = \bx^*} = 
\begin{bmatrix}
\nabla_{\bq} \phi_1^T(\bq^*) s^*
\\
\nabla_{\theta_1} \phi_2^T(\theta_1^*)t^*
\end{bmatrix} = 0
\end{align*}
and the forward part is proved.

Conversely, let $\bx^* = [\bq^*;\theta_1^*]$,  $s^* = \bars(\bx^*)$ and 
$t^* = \bart(\bx^*)$.
Assuming $\nabla_\bx K(\bx)| _{\bx=\bx^*} = 0$, then by \eqref{eq:derivativeK} we have
\begin{align*}
\nabla_{\bq} \phi_1^T(\bq^*)s^* = 0,\quad \nabla_{\theta_1}\phi_2^T(\theta_1^*)t^* = 0
\end{align*}
Under Assumption \ref{assum:pdf}(b), the function $\bq\mapsto \phi_1^T(\bq)s^*$ and $\theta_1\mapsto \phi_2^T(\theta_1)t^*$ has a unique global maximum at $\bar{\bq}(s^*)$ and $\barth_1(t^*)$, respectively. Hence, $\bq^* = \bar{\bq}(s^*)$ and $\theta_1^* = \barth_1(t^*)$ and the converse part is proved. The proof of the Proposition \ref{prop:1} is now complete.
\end{IEEEproof}

\subsection{Identical local maximizer for all nodes}

Let us define $h(s,t) = [h_1(s,t)^T, h_2(s,t)^T]^T$ and $\nabla_{s,t} w(s,t) = [\nabla^T_s w(s,t),\nabla^T_t w(s,t)]^T$.

\begin{proposition}\label{prop:2}
Suppose that Assumptions \ref{assum:pdf} and \ref{assum:lambda} hold, and $w(s,t)$ is continuously differentiable on $\stS \times \stT$. Then, $\langle \nabla_{s,t} w(s,t)  , h(s,t) \rangle  \leq 0$
and $\langle \nabla_{s,t} w(s,t)  , h(s,t) \rangle   = 0 $ if and only if $h(s,t)=0$. 
\end{proposition}
\begin{IEEEproof}
The inner product between $\nabla_{s,t}w(s,t)$ and $h(s,t)$ can be written as
\begin{align}
\bigangle{\nabla_{s,t}w(s,t),h(s,t)}=h_1^T(s,t) \nabla_s w(s,t)+ h_2^T(s,t)\nabla_t w(s,t)
\end{align}
Let $\bar{\bx}(s,t) = [\bar{\bq}^T(s),\barth_1(t)]^T$. Using \eqref{eq:derivativeK} and the chain rule of differentiation,
\begin{align}\label{eq:w_st_1}
\nabla_s w(s,t) &= -\nabla_s \bar{\bx}^T(s,t)
\nabla_\bx K(\bx)|_{\bx= \bar{\bx}(s,t)}\nonumber\\
& = - \nabla_s \bar{\bx}^T(s,t)
\begin{bmatrix}
\nabla_{\bq} \phi_1^T\{\bar{\bq}(s)\} \bars(\bar{\bx}(s,t))
\\
\nabla_{\theta_1} \phi_2^T\{\barth_1(t)\}\bart(\bar{\bx}(s,t))
\end{bmatrix}
\end{align}
Let $l_1(s;\bq) = \phi_{1}^T(\bq)s$ and $l_2(t;\theta_1) = \phi_2 ^T (\theta_1) t$. Note that since $\bar{\bq}(s)$ and $\barth_1(t)$ are the maximum of $\bq\mapsto l_1(s;\bq)$ and $\theta_1 \mapsto l_2(t;\theta_1)$, respectively, we have
\begin{subequations}
\begin{align}
&\nabla_{\bq}\phi_1^T\{\bar{\bq}(s)\}s = 0,\label{eq:l1}\\
&\nabla_{\theta_1}\phi_2^T\{\barth_1(t)\}t = 0.\label{eq:l2}
\end{align}
\end{subequations}
Using definition of $h_1(s,t)$ and $h_2(s,t)$ given in \eqref{eq:h_12_st} and substituting into \eqref{eq:w_st_1}, we obtain
\begin{align}\label{eq:w_st_2}
\nabla_s w(s,t) &= 
-\left[\nabla_s \bar{\bq}^T(s) 
\nabla_{\bq} \phi_1^T\{\bar{\bq}(s)\} h_1(s,t)
+\nabla_s \barth_1^T(t) \nabla_{\theta_1} \phi_2^T\{\barth_1(t)\}h_2(s,t)\right]
\end{align}

Differentiating the function $s\mapsto \Phi_1\{s,\bar{\bq}(s)\}$ where $\Phi_1(s,\bq) \triangleq \nabla_{\bq}l_1(s;\bq)$, we have
\begin{align*}
\nabla_s \Phi_1^T\{s,\bar{\bq}(s)\} = \nabla_s \Phi_1^T(s,\bq)_{\bq = \bar{\bq}(s)}
+\nabla_s \bar{\bq}^T(s)\nabla_{\bq}\Phi_1^T(s,\bq)_{\bq = \bar{\bq}(s)}.
\end{align*}
Since $\nabla_s\Phi_1^T(s,\bq) = \nabla_s(\nabla_{\bq}l_1(s;\bq))^T = (\nabla_{\bq}\phi_1^T(\bq))^T$ and $\nabla_{\bq}\Phi_1^T(s,\bq) = \nabla^2_{\bq}l_1(s;\bq)$, the above equation can be rewritten as 
\begin{align*}
\nabla_s(\nabla_{\bq}l_1(s;\bq)|_{\bq =\bar{\bq}(s)})^T
= (\nabla_{\bq}\phi_1^T\{\bar{\bq}(s)\})^T+ \nabla_s\bar\bq^T(s)\nabla^2_{\bq}l_1(s;\bq)|_{\bq = \bar{\bq}(s)}.
\end{align*}
Because $\nabla_{\bq}l_1(s;\bq)|_{\bq =\bar{\bq}(s)}= 0$, we get 
\begin{align*}
\nabla_{\bq}\phi_1^T\{\bar{\bq}(s)\} = -\nabla^2_{\bq}l_1(s;\bq)|_{\bq = \bar{\bq}(s)} (\nabla_s\bar\bq^T(s))^T.
\end{align*}
Following similar arguments as above, we have
\begin{align*}
\nabla_{\theta_1}\phi_2^T\{\barth_1(t)\} = -\nabla^2_{\theta_1}l_2(t;\theta_1)|_{\theta_1 = \barth_1(t)} (\nabla_t\barth_1^T(t))^T.
\end{align*}
% editing done here
Plugging the above two equations into \eqref{eq:w_st_2} and noting that $\nabla_s\barth_1(t) = \nabla_s t^T \nabla_t \barth_1(t)$, we have
\begin{align*}
h_1^T(s,t) \nabla_s w(s,t) &= h_1^T(s,t) (\nabla_{\bq}\phi_1^T\{\bar{\bq}(s)\})^T\{\nabla^2_{\bq}l_1(s;\bq)|_{\bq = \bar{\bq}(s)}\}^{-1} \nabla_{\bq}\phi_1^T\{\bar{\bq}(s)\} h_1(s,t) \\
&+  h_1^T(s,t) \nabla_s t^T (\nabla_{\theta_1}\phi_2^T\{\barth_1(t)\})^T\{\nabla^2_{\theta_1}l_2(t;\theta_1)|_{\theta_1 = \barth_1(t)}\}^{-1} \nabla_{\theta_1}\phi_2^T\{\barth_1(t)\} h_2(s,t) 
\end{align*}
Using similar arguments as above, we can also show that
\begin{align*}
h_2^T(s,t) \nabla_t w(s,t) &= h_2^T(s,t) \nabla_t s^T (\nabla_{\bq}\phi_1^T\{\bar{\bq}(s)\})^T\{\nabla^2_{\bq}l_1(s;\bq)|_{\bq = \bar{\bq}(s)}\}^{-1} \nabla_{\bq}\phi_1^T\{\bar{\bq}(s)\} h_1(s,t) \\
&+  h_2^T(s,t) (\nabla_{\theta_1}\phi_2^T\{\barth_1(t)\})^T\{\nabla^2_{\theta_1}l_2(t;\theta_1)|_{\theta_1 = \barth_1(t)}\}^{-1} \nabla_{\theta_1}\phi_2^T\{\barth_1(t)\} h_2(s,t) 
\end{align*}
Using the ODE given in \eqref{eq:ode}, we have $h_1^T(s,t) \nabla_s t^T = \dot{s}^T \nabla_s t^T = 2 \dot{t} = 2 h_2^T(s,t)$ and similarly, $h^T_2(s,t) \nabla_t s^T = 2 h_1^T(s,t)$.

Therefore, $\bigangle{\nabla_{s,t}w(s,t), h(s,t)}$ can be expressed as
\begin{align}
&\bigangle{\nabla_{s,t}w(s,t), h(s,t)} \nonumber\\
&= 3h_1^T(s,t) (\nabla_{\bq}\phi_1^T\{\bar{\bq}(s)\})^T\{\nabla^2_{\bq}l_1(s;\bq)|_{\bq = \bar{\bq}(s)}\}^{-1} \nabla_{\bq}\phi_1^T\{\bar{\bq}(s)\} h_1(s,t) \nonumber\\
&+ 3 h_2^T(s,t) (\nabla_{\theta_1}\phi_2^T\{\barth_1(t)\})^T\{\nabla^2_{\theta_1}l_2(t;\theta_1)|_{\theta_1 = \barth_1(t)}\}^{-1} \nabla_{\theta_1}\phi_2^T\{\barth_1(t)\} h_2(s,t) 
\end{align}
where Assumption \ref{assum:pdf}(b) implies that the matrices $\nabla^2_{\bq}l_1(s;\bq)|_{\bq = \bar{\bq}(s)}$ and $\nabla^2_{\theta_1}l_2(t;\theta_1)|_{\theta_1 = \barth_1(t)}$ are negative definite. Therefore, $\bigangle{\nabla_{s,t}w(s,t), h(s,t)}\leq 0$ with equality if and only if \\
${\nabla_{\bq}\phi_1^T\{\bar{\bq}(s)\} h_1(s,t) = 0}$ and ${\nabla_{\theta_1}\phi_2^T\{\barth_1(t)\} h_2(s,t) =0}$.
Assuming $s^*$ and $t^*$ are such that \\
${\nabla_{\bq}\phi_1^T\{\bar{\bq}(s^*)\} h_1(s^*,t^*) = 0}$ and 
${\nabla_{\theta_1}\phi_2^T\{\barth_1(t^*)\} h_2(s^*,t^*) =0}$, or equivalently
\begin{align*}
&\nabla_{\bq}\phi_1^T\{\bar{\bq}(s^*)\} \bars(\bar{\bx}(s^*,t^*)) = \nabla_{\bq}\phi_1^T\{\bar{\bq}(s^*)\} s^* \\
&\nabla_{\theta_1}\phi_2^T\{\barth_1(t^*)\} \bart(\bar{\bx}(s^*,t^*)) = \nabla_{\theta_1}\phi_2^T\{\barth_1(t^*)\} t^*
\end{align*}
By assumption \ref{assum:pdf}, part (b), $\bq^* = \bar{\bq}(s^*)$ is the unique solution to \eqref{eq:l1} and $\theta_1^* = \barth_1(t^*)$ is the unique solution to \eqref{eq:l2}, i.e., $\nabla_{\bq}\phi_1^T\{\bar{\bq}(s^*)\} s^*=0$, $\nabla_{\theta_1}\phi_2^T\{\barth_1(t^*)\} t^* = 0$. Taking into account of the expression of $\nabla_\bx K(\bx)$ given in \eqref{eq:derivativeK}, thus $\nabla_\bx K(\bx)|_{\bx =\bar{\bx}(s^*,t^*)} = 0$. Then, from Proposition \ref{prop:1}, $s^*$ and $t^*$ are roots of $h_1(s,t)$ and $h_2(s,t)$. The proof of Proposition \ref{prop:2} is now complete. 

\end{IEEEproof}

\subsection{The terms $\lambda_n \xi_1^n$, $\lambda_n \xi_2^n$, $\lambda_n r_1^n$ and $\lambda_n r_2^n$ tend to 0 with probability 1 as $n \to \infty$.}

For the discrete time dynamical system \eqref{eq:st_sp_1} to be approximated by the ODE \eqref{eq:ode}, we must show that $\lambda_n \xi_1^n$, $\lambda_n \xi_2^n$, $\lambda_n r_1^n$ and $\lambda_n r_2^n$ asymptotically tend to 0 with probability 1. 

Denoting $M_{1,n,k} \triangleq \sum_{l=n}^k \lambda_l\xi_{1,l}$ and $M_{2,n,k} \triangleq \sum_{l=n}^k \lambda_l\xi_{2,l}$, following the proof of Proposition 1 in \cite{Bianchi2013}, it can be shown that
$M_{1,n,k}$ and $M_{2,n,k}$ are martingales that satisfy \\
$\lim_{n\to\infty} \sup_{k\geq n} \left|M_{1,n,k}\right|=0$ and $\lim_{n\to\infty} \sup_{k\geq n} \left|M_{2,n,k}\right|=0$, with probability 1. In addition, $\lim_{n\to \infty }r_{1,n}=0$ and $\lim_{n\to \infty}r_{2,n}=0$ by Proposition 1 in \cite{Bianchi2013}. Following the proof of Theorem 1 in \cite{Cappe2009}, it can then be shown that $\lim_{n\to \infty} d(\bx_i^n,\stE)=0$ with probability 1, which completes the proof of Theorem~\ref{th:conv_max}.

\section{TDOA measurement setup}
\label{sec:tdoa_meas}

\subsection{TDOA measurement principle}
\label{subsec:tdoa_princ}

Let us start by emphasizing that our TDOA measurement setup \textit{does not} aim at reproducing the distributed EM algorithm, but is used to measure the TDOA between different pair of nodes so that the distributed EM algorithm can be evaluated over real measurements offline. 

\medskip

Even when synchronized with GPS, a time offset remains between the different USRP receive nodes. When instructed to start measuring at a given time $T_0$, each node $i$ will in reality start measuring at time $T_{0i}$ which differs slightly from time $T_0$. This time offset between nodes results in a TDOA error that can be as large as 60~m. We now introduce a relaying architecture that cancels out this remaining time offset, provided that the clock skew is close to zero (which is the case in practice when using GPS-disciplined local oscillators). Let us consider the signal $x(t)$ sent by the transmitter. If we omit the effects of fading and noise for readability, the signal at the $i$-th node is given by $r_i(t)=x(t-\tau_{i1})$ where $\tau_{i1}=d_i/c_0$ is the propagation delay between the transmitter and node $i$ ($c_0$ is the speed of light). The relay will start sampling the message at time $T_{0i}$ (which might differ slightly for different relays), resulting in the sampled message $r_i[k] = x(T_{0i}+ kT_s -\tau_{i1})$. Each relay node $i$ will then forward its received message to a central receiver after a certain time delay $T_{Di}$, where a different time slot is allocated to each relay as to avoid collisions. The transmitted message from each relay is then given by
\begin{align*}
t_{i}(t) = \sum\limits_{l=-\infty}^{\infty} x(T_{0i}+lT_s-\tau_{i1}) \cdot u(t-lT_s-T_{0i}-T_{Di})
\end{align*}
where $T_{Di}$ is the retransmission delay of the $i$-th node, and $u(t)$ is the pulse shaping filter of the node. The central receiver will start sampling the message from relay $i$ at time $T_{0R}+ T_{Di}$, where $T_{0R}$ is the central receiver's estimate of $T_0$. The sampled received message at the final receiver can be written as
\begin{multline*}
r_R^{(i)}[n] = \sum\limits_{l=-\infty}^{\infty} x(T_{0i}+lT_s-\tau_{i1}) \cdot g( T_{0R}+T_{Di} + nT_s- lT_s-T_{0i}- T_{Di} - \tau_{i2})
\end{multline*}
where $\tau_{i2}$ is the propagation delay between the $i$-th relay and the final receiver, and $g(t)=u(t)\ast u'(t)$ with $u'(t)$ being the pulse shaping filter at the receiver. If the pulse shaping filters are chosen appropriately and inter-symbol interference is canceled, the previous equation can be simplified to 
\begin{multline}
r_R^{(i)}[n] = \sum\limits_{l=-\infty}^{\infty} x(T_{0i}+ lT_s-\tau_{i1}) \cdot \delta( T_{0R}+ T_{Di} + (n-l)T_s-T_{0i}- T_{Di} - \tau_{i2})
	\label{eq:recv_mess_1b}
\end{multline}
where $\delta(t)$ is the Dirac function. In \eqref{eq:recv_mess_1b}, only one term in the sum is non-zero for a given $n$. In that case \eqref{eq:recv_mess_1b} simplifies to
\begin{equation*}
	r_R^{(i)}[n] = x( T_{0R} + nT_s  - \tau_{i2} - \tau_{i1}  )
\end{equation*}
which is independent of the node measurement time $T_{0i}$. For different nodes (which forward the messages with different delays $T_{Di}$), the relayed messages at the central receiver will have identical offset $T_{0R}$. The proposed architecture is thus successfully able to cancel out differences in node measurement time offsets. The receiver then computes the ambiguity functions between the received messages $r_R^{(i)}[n]$ and $r_R^{(j)}[n]$ from nodes $i$ and $j$. The index of the peak of the ambiguity function is then equal to $\tau_{i2}+\tau_{i1}-\tau_{j2}-\tau_{j1}$. If $\tau_{i2}$ and $\tau_{j2}$ are known, the receiver can recover the original TDOA $\tau_{i1}-\tau_{j1}$. The receiver is then able to compute the TDOA (expressed in distance) with respect to node 1 as follows: 
\begin{equation*}
\td_{i1} = \left(\tau_{i1}-\tau_{11}\right)\cdot c_0
\end{equation*}

\subsection{Experimental setup details}
\label{subsec:exper_details}

In our experimental setup, the relay nodes use USPR-N210 with WBX boards, which allows them to receive the signal from the transmitter on frequency $f_1$, and transmit the message after a delay $T_{Di}$ to the central receiver over frequency $f_2$. The USRP drivers allow for very fine time control (down to one sample) of the received and transmitted samples. The central receiver consists of a USRP-N210 with a TVRX2 board, which is able to receive the signal from the transmitter on frequency $f_1$ and the signals from the relays on frequency $f_2$. All the USRPs in our setup use GPSDO modules, which allows them to synchronize their internal clocks and local oscillators to the GPS UTC time. This is precise enough to have quasi-zero clock skew, but the clock offset between nodes can still be as high as 100~ns. In a off-line post-processing step, the central receiver oversamples the received messages by a factor of 10 and passes these signal through a low-pass filter, which allows to increase the resolution accuracy of our setup to 10~ns. Our setup was calibrated and tested extensively in a controlled environment with cables to avoid multipath and limited SNR problems. In those cases, our setup has an error below 10~ns. In an outdoor LOS setting, the TDOA error of our setup has a mean of 1~m and a standard deviation of 4~m. The RF source was a signal generator sending a random D-BPSK sequence. All the parameters of the setup are given in Table~\ref{tab:system_param}. During the experiments, node S1 was chosen as the receiver node, while S0, S2 and S3 act as relay nodes. 
\begin{table}
	\centering
	\caption{Experimental setup system parameters}
		\begin{tabular}{lc}
			\hline
														\textbf{Parameter} 							& Value 			\\
			\hline 									
			 											RF source frequency	$f_1$				& 795~MHz 	\\
			\rowcolor{gray90}			RF source signal bandwidth 		  & 1~MHz 	\\			
														Relay/Rx sample rate						& 10~MHz 	\\
			\rowcolor{gray90} 		Relaying channel frequency $f_2$& 755~MHz 	\\
														Relay delay $T_{Di}$ 						& $\{25,50,75\}$~ms 		\\	
			\rowcolor{gray90} 		Recorded packet length					& 10~ms 	\\																		
			\hline
		\end{tabular}
	\label{tab:system_param}
\end{table}

\medskip

In our TDOA measurement setup, one needs to know the propagation delay between the relay node and the central receiver to be able to estimate the propagation delay between the transmitter and the relay node. In our setup, this is achieved by creating a LOS link between the relay transmit antenna and the receiver. This is done by placing all the USRPs in Figure \ref{fig:picture_meas} on the same floor as the transmitter. The NLOS link for nodes $S_2$, $S_3$ and $S_4$ is created by using a 15~m cable between the USRP and the antenna that receives the signal over frequency $f_1$, which allows us to place this antenna on the upper floor corresponding to $S_2$ and $S_4$ in Figure~\ref{fig:picture_meas}, and behind the corner in the case of $S_3$. All the other antennas (for transmitting the relayed message in the case of $S_1$, $S_3$ and $S_4$, for receiving the relayed message in the case of $S_2$) are placed on the lower floor, and have a LOS to each other. The propagation delay between relay and receiver is then evaluated by measuring the distance between relays and receiver.

\section*{Acknowledgements}
The authors would like to thank Mr~Cheng~Chi and Dr~Zahra~Madadi for their help with the experimental measurements. 

\footnotesize{
\bibliographystyle{IEEEtran}
\bibliography{IEEEabrv,refs_distr_em,biblio}
}

\end{document}